\documentclass{article}
\usepackage{arxiv}

\usepackage[utf8]{inputenc} 
\usepackage[T1]{fontenc}    
\usepackage{url}            
\usepackage{booktabs}       
\usepackage{amsfonts}       
\usepackage{nicefrac}       
\usepackage{microtype}      
\usepackage{authblk}        
\usepackage{amsmath}
\usepackage{lipsum}
\usepackage{siunitx}
\usepackage[backend=biber,
			natbib=true,
			style=numeric,
			sorting=none,
			isbn=false,
			url=false, 
			doi=false,
			eprint=false]{biblatex}

\AtEveryBibitem{\clearfield{month}}
\usepackage{xfrac}
\usepackage{amssymb}
\usepackage{xr-hyper}
\usepackage{hyperref}
\usepackage{etoolbox}
\newtoggle{keywords}
\toggletrue{keywords}
\newtoggle{abstractBeforeMakeTitle}
\togglefalse{abstractBeforeMakeTitle}
\graphicspath{{figures/}}
\newtoggle{usebiber}
\toggletrue{usebiber}
\newlength{\hcolw}
\newlength{\hdoublecolw}
\begin{document}

\newcommand{\Smatrix}{$\mathbf{\hat{S}}$}
\newcommand{\Soneone}{$\mathbf{\hat{S}_{11}}$}
\newcommand{\Sonetwo}{$\mathbf{\hat{S}_{12}}$}
\newcommand{\Stwoone}{$\mathbf{\hat{S}_{21}}$}
\newcommand{\Stwotwo}{$\mathbf{\hat{S}_{22}}$}
\newcommand{\Snm}{$\mathbf{\hat{S}_{nm}}$}
\newcommand{\kvector}{$\vec{k}$}
\newcommand{\ktrans}{$\vec{k}_{\perp}$}
\newcommand{\sapphire}{Al$_2$O$_3$}

\title{Nanopatterned Sapphire Substrates in Deep-UV LEDs: Is there an Optical Benefit?}
\author[1,3]{Phillip Manley*}
\author[2]{Sebastian Walde}
\author[2]{Sylvia Hagedorn}
\author[4]{Martin Hammerschmidt}
\author[3,4]{Sven Burger}
\author[1]{Christiane Becker}
\affil[1]{Helmholtz Zentrum Berlin f{\"u}r Materialien und Energie, Berlin, Germany}
\affil[2]{Ferdinand-Braun-Institute, Leibniz-Institut f{\"u}r H{\"o}chstfrequenztechnik,  Berlin, Germany}
\affil[3]{Zuse Institute Berlin, Berlin, Germany}
\affil[4]{JCMwave, Berlin, Germany}
\affil[*]{Corresponding author: phillip.manley@helmholtz-berlin.de}




\iftoggle{abstractBeforeMakeTitle}
{
}
{
	\maketitle
}

\begin{abstract}
	Light emitting diodes (LEDs) in the deep ultra-violet (DUV) offer new perspectives for multiple applications ranging from 3D printing to sterilization. However, insufficient light extraction severely limits their efficiency. Nanostructured sapphire substrates in aluminum nitride based LED devices have recently shown to improve crystal growth properties, while their impact on light extraction has not been fully verified. We present a model for understanding the impact of nanostructures on the light extraction capability of DUV-LEDs. The model assumes an isotropic light source in the semiconductor layer stack and combines rigorously computed scattering matrices with a multilayer solver. We find that the optical benefit of using a nanopatterned as opposed to a planar sapphire substrate to be negligible, if parasitic absorption in the p-side of the LED is dominant. If losses in the p-side are reduced to 20\%, then for a wavelength of 265\,nm  an increase of light extraction efficiency from 7.8\% to 25.0\% is possible due to nanostructuring. We introduce a concept using a diffuse ('Lambertian') reflector as p-contact, further increasing the light extraction efficiency to 34.2\%. The results underline that transparent p-sides and reflective p-contacts in DUV-LEDs are indispensable for enhanced light extraction regardless of the interface texture between semiconductor and sapphire substrate. The optical design guidelines presented in this study will accelerate the development of high-efficiency DUV-LEDs, but the model is also readily applicable to other multilayer opto-electronic nanostructured devices such as photovoltaics or photodetectors.
\end{abstract}

\iftoggle{keywords}{
}
{
}

\iftoggle{abstractBeforeMakeTitle}
{
	\maketitle
}
{
}

\section{Introduction}
Light emitting diodes (LEDs) have become a key technology on the march towards a sustainable society. Lately, the push towards deep ultra-violet (DUV) LED devices has unlocked further applications such as 3D printing \cite{Becker2015,Hollander2018}, optical storage \cite{Riesen2018}, water treatment \cite{Jones2014} and sterilization \cite{Yin2013}. Currently the leading technology for DUV LEDs is based on III-nitride materials .

The problem of light extraction is common to all LED devices \cite{Saxena2009,Gessmann2004}. Light is generated inside a material which has a higher refractive index than its surroundings. Due to the barrier of total internal reflection, a significant fraction of the light created will be trapped inside the LED. This requires a suitable strategy for enhancement of the light extraction efficiency (LEE) to overcome this limitation. Some of the strategies for increasing LEE that have been employed for LED devices include rough surfaces \cite{Fujii2004,Windisch1999},  micro-lenses \cite{Greiner2007,Madigan2000} , plasmonic gratings \cite{Zhang2013}, dielectric nanoparticles \cite{Lee2014,Mont2008}, bio-mimetic structures \cite{Pernot2010}, nanophotonic structures  \cite{Inoue2017} and photonic crystals \cite{Oder2003,Kim2005,Matioli2010}.

\begin{figure*}[htbp]
	\centering
	\includegraphics[width=0.8\hdoublecolw]{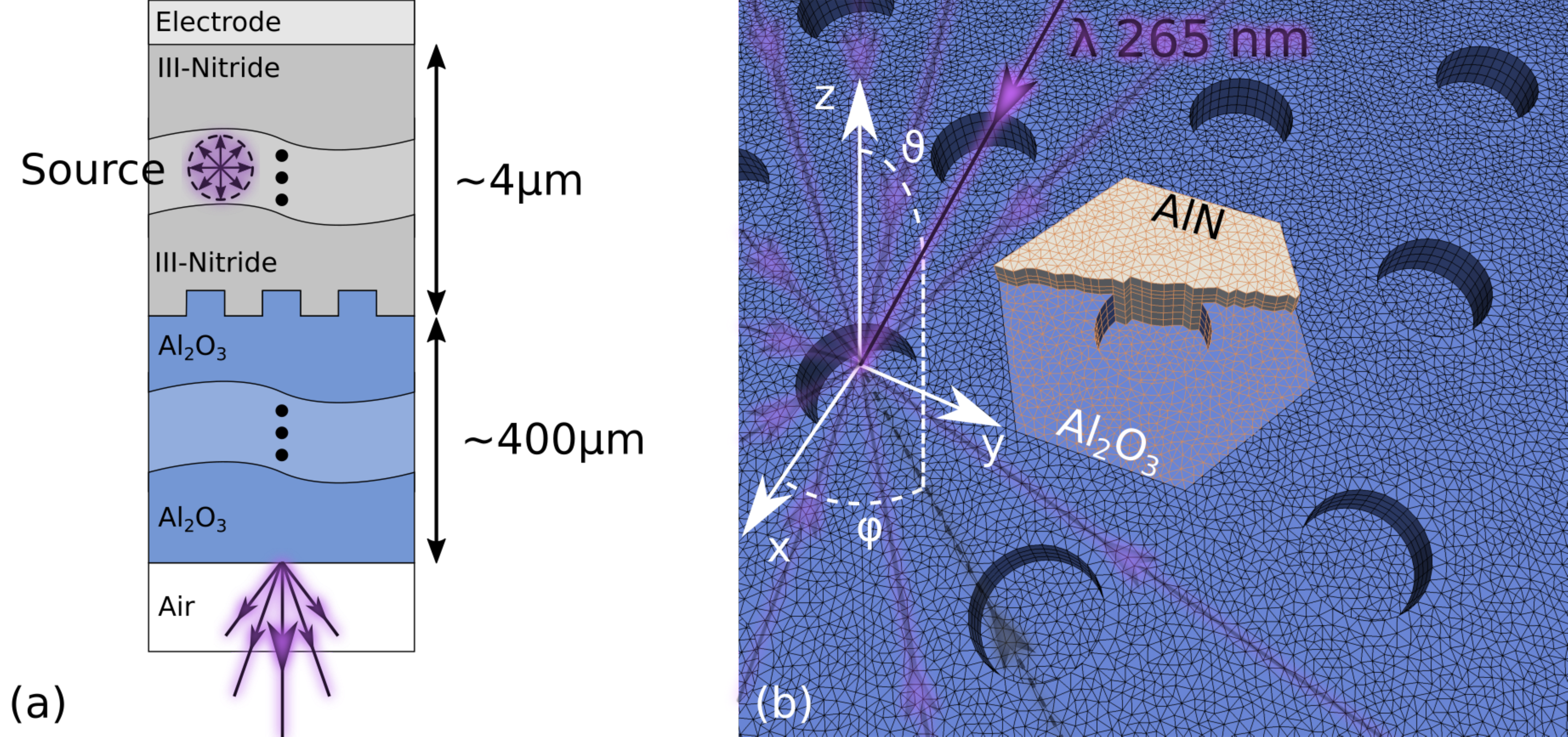}
	\caption{(a) Simplified sketch of the III-nitride LED device. The thin film LED contains the light source and sits on a {\sapphire} substrate. Light is emitted through the substrate into air. The interface between III-nitride and {\sapphire} has been nanopatterned. (b) The nanopatterned {\sapphire} substrate with a triangular array of nanoholes. The periodic unit cell used for simulations is highlighted, the upper AlN layer (part of the III-nitride device) which completely fills the nanoholes has been partially hidden for visibility. UV light with 265 nm vacuum wavelength is incident from above the surface due to emission inside the III-nitride device. Light is incident to the surface with a polar angle $\theta$ and an azimuthal angle $\phi$.}
	\label{fig:MultiGrid}
\end{figure*}

Any light extraction strategy used must be compatible with the device structure. Figure \ref{fig:MultiGrid}(a) shows a simplified schematic of the LED device. The thin film multilayer system which constitutes the LED, here represented as a single III-nitride layer, has a thickness of around 4\,$\mu$m. Whereas, the {\sapphire} (sapphire) substrate has a thickness on the order of 400\,$\mu$m. Concepts such as micro-lenses, rough surfaces and nanoparticles may be of use at the {\sapphire}/air interface where they are well separated from the LED thin film. However, they are less applicable at the AlN/{\sapphire} interface. The relatively large size of micro-lenses makes them incompatible with the thin film LED. Rough surfaces and nanoparticles are likely to induce defects in crystal growth. Although the plasma frequency of metals such as Al, Mg, Ga and Rh is compatible with DUV applications \cite{Gutierrez2018}, introducing these materials to the AlN/{\sapphire} interface may also create a source for defects.

Photonic crystal structures may offer optical benefits while being highly compatible and even beneficial to standard processing techniques. III-nitride based LEDs are typically grown on {\sapphire} substrates as they are commercially available with low optical loss in the DUV wavelength range. However, this has the drawback of a lattice mismatch between {\sapphire} and the III-nitride LED heterostructure which can lead to a large number of threading dislocations \cite{Ponce1997}. Structuring the {\sapphire} surface can lead to an alleviation of stress at the interface, thereby improving the material quality of the LED layers in the case of sub-micometer structures on the {\sapphire} surface \cite{Kim2011,Dong2013,Lee2017}. This kind of structured {\sapphire} substrate is more commonly referred to as a nanopatterned sapphire substrate (NPSS). Takano et al. recently realized DUV-LEDs on NPSS with external quantum efficiencies larger than 20\% \cite{Takano2017}. Figure \ref{fig:MultiGrid}(b) shows a top down view of a simulation mesh of such a NPSS based on nanohole array in a triangular lattice.

Based on experience with visible LEDs it is assumed that the structured interface between {\sapphire} and the DUV-LED will increase the LEE. However, recent results indicate doubts at the universal validity of NPSS causing an optical benefit \cite{Nagasawa2018}. Even in the above mentioned work by Takano et al. NPSS is only applied as an additional feature in combination with other light extraction strategies, but not alone \cite{Takano2017}. Hence, while photonic crystals may be optically promising, since they offer the ability to extract modes trapped in both the substrate and the LED layer stack, the specific conditions under which their use is advantageous in DUV-LEDs need to be carefully analyzed.

In order to perform this analysis, a simulation framework for obtaining the optical characteristics of a layered system with structured interfaces and, in particular, an isotropic (omni-directional) light source needs to be established. Three features  present in the LED system will help determine the type of model appropriate. (1) Broad range of emission angles inside the device, (2) Low index contrast between layers inside the III-nitride thin film, and (3) Coherence length of light emission $\leq$ III-nitride device thickness. In the following we discuss how these three features impact the choice of model. 

The exact angular and polarization dependence of light emission is heavily dependent on the ratio of vertical to lateral dipoles in the multiple quantum well (MQW) emission layer. We assume an \textbf{isotropic distribution of emission angles (1)} and equal weighting of polarisations. The LED device is made up of multiple different layers. However, the majority of layers lie in the Al$_{x}$Ga$_{1-x}$N system. The refractive index contrast between GaN and AlN at 265\,nm wavelength is only ~0.2, meaning that the majority of layers will have a very small refractive index contrast. Furthermore, the refractive index is often graded between layers of different compositions, thus further reducing the index contrast seen by light. Thus, we approximate the entire LED device as a \textbf{homogeneous AlN layer (2)}. Finally, the LED device has a thickness roughly 3-4\,$\mu$m, which is approximately equivalent to the longitudinal coherence length of 265\,nm light emitted in AlN with a Gaussian line shape and 10\,nm full width at half maximum \cite{Akcay2002}. Therefore, multiple passes through the LED device will lose coherency with each other, meaning that \textbf{coherence effects can be neglected (3)}. The sapphire substrate, being hundreds of $\mu$m thick, should also be modelled incoherently.

With these model assumptions, a natural method to describe this system is that of scattering matrices. Scattering matrix methods have been employed for multilayer systems in many areas of physics including solid state physics \cite{Ko1988}, acoustics \cite{Reinhardt2003} and optics \cite{Whittaker1999,Li1996}. In its simplest form, the scattering matrix method couples homogeneous layers together, meaning that light can be described as plane waves inside each layer.

In order to model the scattering at the interfaces, many different approaches such as rigorous coupled wave analysis (RCWA) \cite{Moharam1981}, finite difference time domain (FDTD) \cite{Yee1966} or the finite element method (FEM) \cite{Lavrinenko2014} can be used.

Figure \ref{fig:MultiGrid}(b) shows a mesh for a geometry consisting of circular inclusions in the $x-y$ plane. Considering e.g. nanofrustums, it will also be necessary to model structures which are inhomogeneous also in the $z$ direction. Methods such as RCWA and FDTD which use structured meshes often resort to a staircasing to approximate such geometries. These shapes are better approximated with unstructured grids such as those used in the finite element method (FEM). The convergence properties of FEM have also been shown to be favourable compared to other methods \cite{Burger2005}. This is important for obtaining scattering matrices, since the coupling of light to a very high number of diffraction orders is required, which needs a highly accurate solution.

Previous work on modelling the LEE of DUV-LEDs has employed FDTD \cite{Ooi2018,Lee2017,Ryu2013}. In this case the entire device was modelled coherently. This meant reducing the thicknesses of layers, in particular the {\sapphire} substrate, in order to make the simulation tractable. We instead choose to simulate the interfaces using FEM and subsequently couple them together in a scattering matrix stack system. This significantly reduces the computational complexity and can remove coherent effects which may not be applicable to real devices.

A significant source of optical loss in DUV-LEDs is that of absorption in the highly absorbing p-side. Due to this there have been reports in the literature of low absorption alternatives \cite{Martens2016,Liang2018,Hirayama2014}. In order to further understand the role of the NPSS in LEE, we combine the NPSS with two kinds of rear reflector. Typical III-nitride LEDs have planar p-contact electrodes which will act as a mirror for incident light, with reflectance dependent on the specific metal used. In addition to this kind of a flat electrode, we also investigate the effect of a diffuse (Lambertian) type reflector. Structures which induce diffuse scattering have been employed in LEDs and solar cells to enhance both light extraction \cite{Saxena2009,Cho2014} and light trapping \cite{Goetzberger1981,Yablonovitch1982}. This could be realised by combining a structured transparent layer that provides the necessary diffuse reflection with a rear planar metallic electrode, in order to avoid plasmonic losses induced in structuring the metal. A concept using a photonic crystal embedded in a transparent p contact as a rear reflector has been presented \cite{Kashima2017}. However, photonic crystals are only highly reflective for a narrow ranges of angles. A diffuse scatterer which is angular independent is preferable for extracting all of the light trapped in the LED.

Previous work analysing the impact of NPSS on the LEE in DUV-LEDs has focused on single structures \cite{Ryu2013,Hirayama2014,Lee2017} or a range of dimensions for a single geometry \cite{Ooi2018}. In this work we analyse a range of nanostructure dimensions, as well as three different kinds of geometry. 

In the following sections we firstly develop a robust simulation framework for analyzing the optical characteristics of structured interfaces (section \ref{sec:Theory}). This involves a method to efficiently determine scattering matrices for periodic nanostructures. The scattering matrices are then combined to obtain the response of the entire multi-layer stack. We then apply this simulation framework to the case of a structured sapphire – AlN interface and discuss the results in the context of LEE in DUV LEDs (section \ref{sec:ResultsAndDiscussion}). The crucial AlN/{\sapphire} interface is analysed first in isolation from other interfaces in the LED. In particular the effect of geometry and pitch on the transmittance is shown. Building on these results the contribution to light extraction of a rear reflector, either specular or diffuse, is quantified. As a a final step the full device stack is simulated to estimate the light extraction efficiency for the most promising designs proposed here.

\section{Method}
\label{sec:Theory}
The simulation framework used for the analysis in section \ref{sec:ResultsAndDiscussion} is comprised of calculating scattering matrices for the material interfaces and combining said matrices into a system of equations to describe light propagation in the multi-layer device. For more information on the method please consult the supplementary material.

The scattering matrix ({\Smatrix}) describes how light is reflected and transmitted at a material interface. The entries for each $i,j$ of {\Smatrix} define the coupling between modes in two different material layers. If the interface between layers happens to be planar then the coefficients are simply the Fresnel coefficients. For non-planar interfaces, numerical approaches may be employed to obtain the coupling coefficients. In the present work we use the finite element method solver JCMsuite \cite{Pomplun2007} in order to obtain the coupling coefficients.

There are many examples of multi-layer solvers in the literature \cite{Liu2012,Eisenlohr2015,Santbergen2017}. Many such solvers work iteratively in which case the matrix operations are applied to a source vector repeatedly until the amount of energy remaining in the system goes below a certain threshold. In this work we use the incoming and outgoing fluxes in each layer to form a system of linear equations connected via the scattering matrices. This system of equations can be solved to obtain the incident and outgoing flux at each interface \cite{Hammerschmidt2016}.

The fluxes of modes outgoing from the stack system represent the transmittance, i.e. light extracted from the device. For each mode in the source layer, the sum of all transmittance modes represents the transmittance when that mode is illuminated,

The total transmittance can be obtained by numerically integrating the transmittance over the hemisphere formed from the angular distribution of modes in the source medium, normalised to the area of the hemisphere.
\begin{equation}
T = \frac{1}{2\pi}\int^{2\pi}_{0}\int^{\pi/2}_{0} T(\theta,\phi) \sin\theta d\theta~d\phi.
\label{eq:AngularIntegration}
\end{equation}
Where $\theta$ and $\phi$ represent the polar and azimuthal angles, respectively.

\section{Results and Discussion}
\label{sec:ResultsAndDiscussion}
\subsection{Single Pass Transmittance}
\begin{figure*}[htbp]
	\centering
	\includegraphics[width=\hdoublecolw]{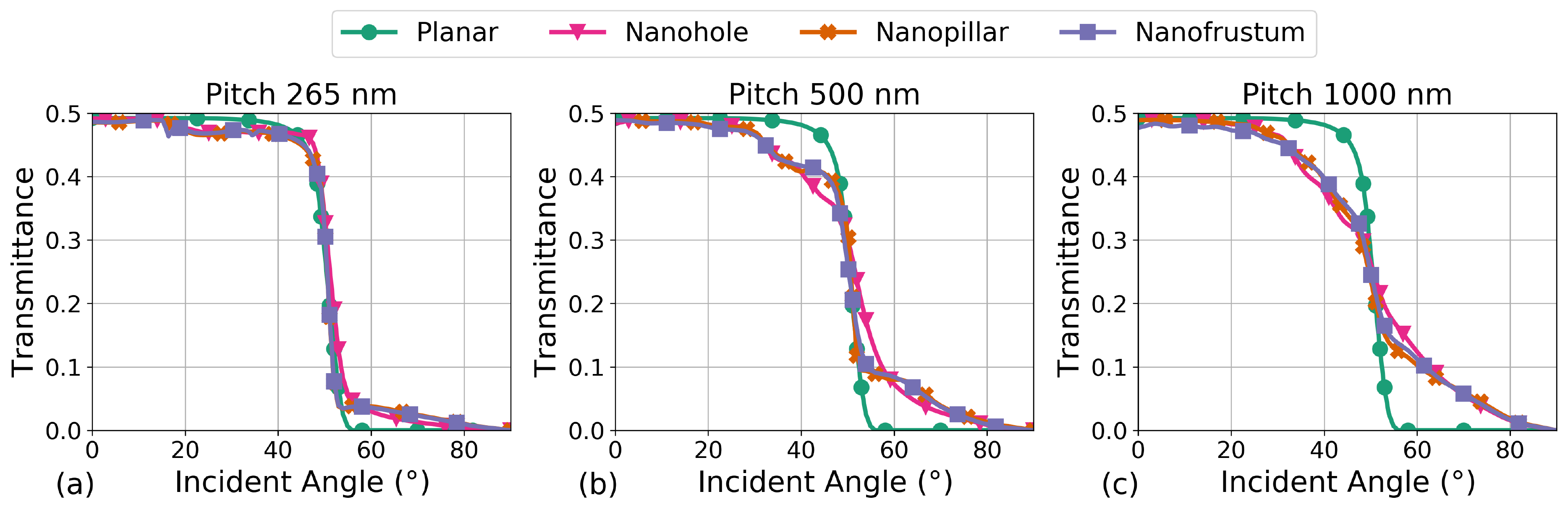}
	\caption{Angular dependence of the single pass transmittance for various nanostructures and a planar interface between AlN and sapphire. The values of transmittance represent the polarization and azimuthal angle averages. The pitch of the nanostructure  array is 265 nm (a), 500 nm (b) and 1000 nm (c). Due to the absorption of all light initially incident in the $+z$ direction, the maximum value for transmittance is 0.5.}
	\label{fig:SinglePassTransmission}
\end{figure*}

\begin{figure}[htbp]
	\centering
	\includegraphics[width=0.6\hcolw]{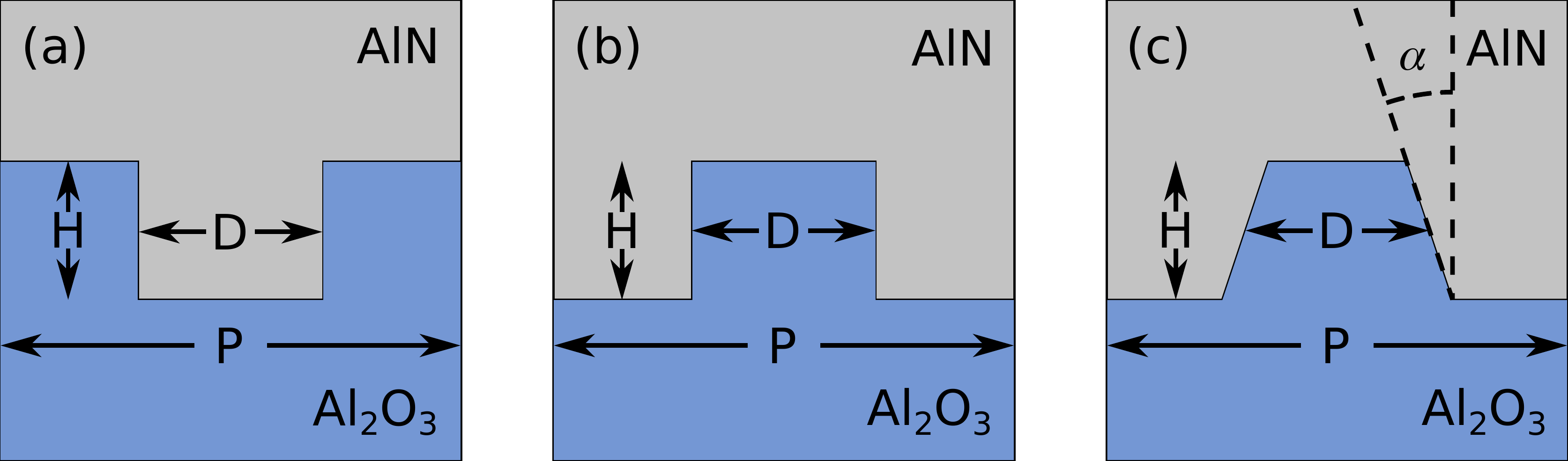}
	\caption{A schematic the cross sectional geometry of the three nanostructures presented in the analysis (a) nanoholes, (b) nanopillars and (c) nanofrustums.}
	\label{fig:geometries}    
\end{figure}

In order to assess the suitability of a nanopatterned sapphire substrate for light extraction enhancement, we initially consider the angular response of the nanostructured interface between AlN and {\sapphire} substrate in isolation from the entire LED device. Three different kinds of nanopattern have been investigated, each consisting of an axially symmetric shape at the center of a triangular lattice with pitch $P$. An example of the nanohole lattice as seen from above is shown in figure \ref{fig:MultiGrid}. Schematic images of the cross section of each geometry is shown in figure \ref{fig:geometries}. The nanohole array (a) consists of cylindrical inclusions of AlN inside the {\sapphire}, while the nanopillar array (b) consists of cylindrical inclusions of {\sapphire} in AlN. The nanofrustum array (c) is equivalent to the nanopillar array except for an angular tilt to the side walls ($\alpha$) of the pillar. For the simulations presented here $\alpha$ = 20\si{\degree} was chosen as a close approximation to realistic interface geometries \cite{Hagedorn2016,Zhang2016}. In each of the array types the diameter $D$ (frustums: diameter at half height) of the inclusion was set to $D = P/2$ and the height $H$ of the inclusion was set to $H = P/4$. 

\begin{table}[htbp]
	\centering
	\begin{tabular}{c|c|c|c|c}
		Pitch (nm) & Planar & Holes & Pillars & Frustums \\
		\hline
		265 & 17.6\% & 18.6\% & 18.3\% & 18.3\% \\ 
		500 & 17.6\% & 18.6\% & 18.7\% & 18.6\% \\
		1000 & 17.6\% & 19.2\% & 18.9\% & 19.0\% \\
	\end{tabular}
	\caption{The integrated single pass transmittance for the curves presented in figure \ref{fig:SinglePassTransmission}.}
	\label{tab:SinglePassTable}
\end{table}

Figure \ref{fig:SinglePassTransmission} presents the polarization averaged transmittance as a function of incident angle for light at an interface between AlN and sapphire, incident from the AlN side with 265 nm vacuum wavelength. The azimuthal angular dependence has been numerically integrated out in order to focus on the polar angle dependence. 

This gives an estimate of the maximum light extraction for the LED considering a highly absorbing p-side above the MQW light source. In that case, emission from the MQWs in the $+z$ direction will be totally absorbed and does not contribute to transmittance into the substrate. This limits the transmittance to a maximum value of 0.5. Emission in the $-z$ direction down towards the AlN/sapphire substrate will contribute only once, since any light reflected at this interface will be absorbed inside the highly absorbing upper side. 

The planar interface shows relatively high transmittance for small incident angles, but rapidly drops to zero close to the critical angle of 51\si{\degree}. Using \ref{eq:AngularIntegration} and assuming an isotropic light source this results in an angular integrated total transmittance of 17.6\% for the planar case. The integrated transmittance for each of the curves in figure \ref{fig:SinglePassTransmission} is shown in table \ref{tab:SinglePassTable}. Each of the nanostructures has the tendency to flatten out the transmittance curve, meaning that the small angle transmittance is lower than for the planar case, while the large angle transmittance above the critical angle becomes non-zero.

The trend visible in figure \ref{fig:SinglePassTransmission} is that the various nanostructures themselves do not differ strongly from each other for a given value of pitch. Increases in the pitch lead to further flattening of the transmittance curve. This suggests that the scattering is largely determined by the lattice pitch and is rather insensitive to the particular shape of scatter inside the periodic unit cell. The increase of scattering for larger pitches can be understood from the fact that the larger pitch will result in shorter reciprocal lattice vectors, thereby increasing the number of diffraction orders available. The larger number of diffraction orders provides more pathways for light to be scattered, allowing the transmittance to flatten compared to the planar case (where no scattering pathways are present).

The net effect is a minimal increase in the total transmittance, mainly due to the fact that more energy is present in the higher emission angles. Therefore, an increase in transmittance at high angles more than compensates for a comparable decrease in transmittance at smaller incident angles. The overall increase in transmittance by nanotexturing is relatively small, the maximum being 9\% relative for 1000 nm pitch nanoholes. This suggests that the NPSS alone will not have a large influence on increasing the LEE of DUV-LEDs. However, if the light was able to have multiple chances at transmittance, the NPPS would significantly increase the transmittance, since they present transmittance values larger than zero for the whole angular range, unlike the planar case which is limited via total internal reflection.

\subsection{Multiple Pass Transmittance}
\begin{figure}[htbp]
	\centering
	\includegraphics[width=0.6\hcolw]{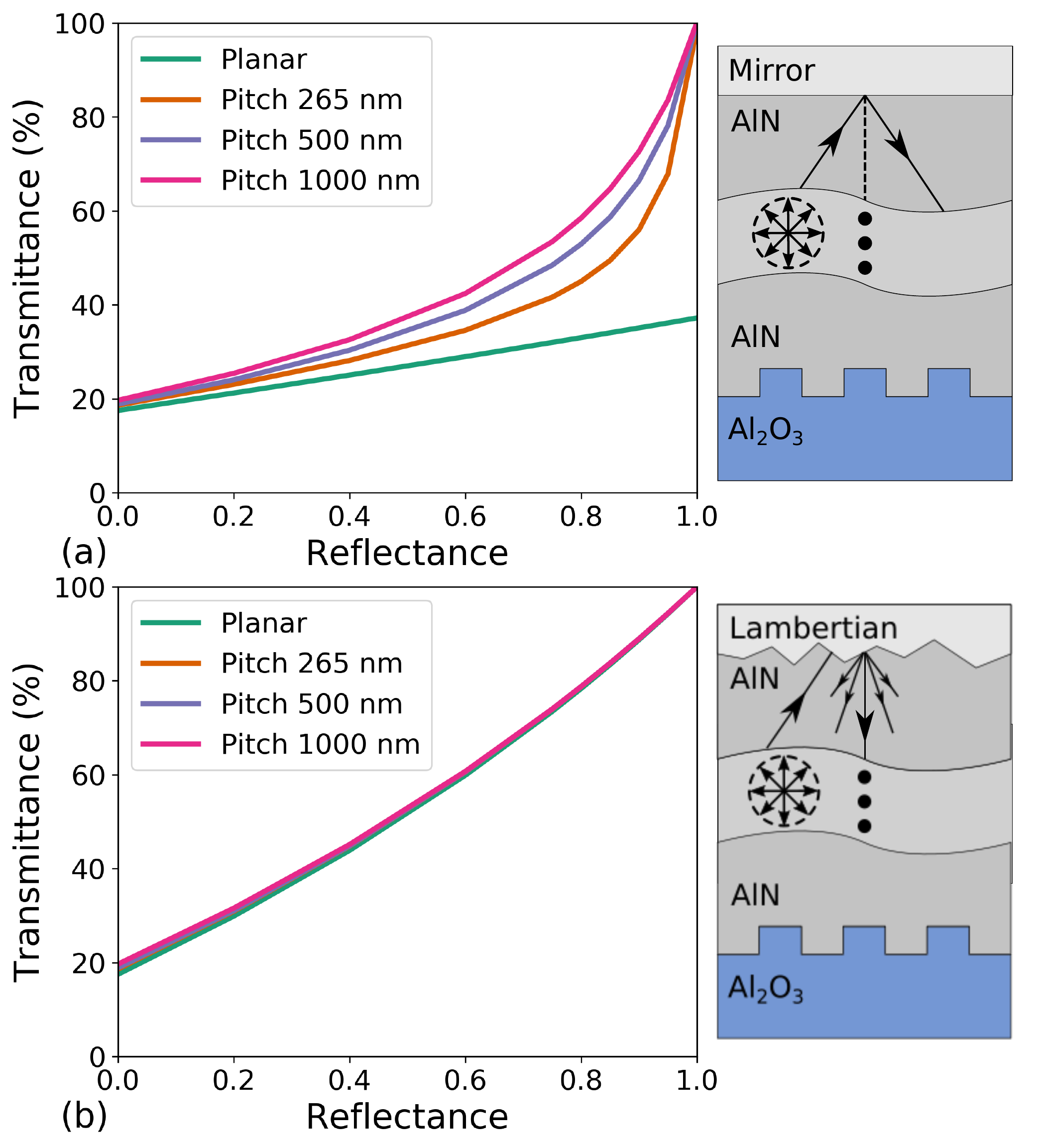}
	\caption{Dependence on rear reflector reflectance of the total transmittance into sapphire for a planar interface as well as nanohole arrays of three different pitches. In (a) the rear reflector is a mirror while in (b) it is a diffuse reflector. The inset in each case shows a schematic image of the layered system under consideration. The source is between two interfaces, which are defined via scattering matrices.}
	\label{fig:MultiPassTransmission}
\end{figure}

While nitride based DUV-LEDs typically have a highly absorbing p-side above the MQWs, efforts towards a transparent p-side have been reported in the literature \cite{Hirayama2014,Khan2019}. In this section we analyze the optical benefit of NPSS on the LEE from the AlN layer into the sapphire substrate, considering a transparent p-side above the MQWs and a reflective p-contact.

In order to inject carriers a metallic electrode layer is typically deposited on the p-side as a p-contact. Optically a planar metallic layer will act as a mirror. There have been multiple reports of different contact materials in the literature with attempts to increase the reflectance of the rear contact while maintaining good electrical properties \cite{Takano2017,Khan2019}. Instead of evaluating the effect of the rear mirror for any one particular material, we instead assume a mirror with an angular independent reflectance $R$ value that can be varied in order to understand the effect of a rear reflector. Different values of $R$ give different estimates of the losses present both in the p-side and p-contact reflector in a real device.

While a mirror reflector will conserve the angular distribution of incident light, other reflective surfaces may not. Diffuse scattering surfaces, otherwise known as Lambertian scatterers, will redistribute the angular distribution of incident light such that the intensity of reflected light $R(\theta) = R_{0}\cos\theta$, with $R_{0}$ chosen such that the total intensity reflected from the surface is given by $R$. With this definition the mirror (specular) and diffuse reflector can be compared with the same overall reflectance.

Figure \ref{fig:MultiPassTransmission} shows how the total transmittance from AlN into the {\sapphire} substrate varies with reflectance of the rear reflector for the specular (a) and diffuse (b) case. First considering the specular reflector, it is apparent that the transmittance for a planar interface increases almost linearly with the rear reflectance, varying from 17.6\% to 37.22\%. The maximum value is limited due to total internal reflection, even with unity reflectance at the rear side. This is in contrast to the nanohole arrays which have a highly nonlinear increase in total transmittance with rear reflectance, ultimately culminating in 100\% transmittance when the rear reflectance is 1. This is due to the non-zero transmittance above the critical angle induced via diffraction. The different increase in the total transmittance for the different pitches can be understood via the curves in figure \ref{fig:SinglePassTransmission}. There it is shown that the high angle transmittance increases as the nanostructure pitch increases. Likewise in figure \ref{fig:MultiPassTransmission} the rate at which the total transmittance reaches 100\% increases for increasing pitch, emphasising the need for extraction of the high angle modes. Assuming a realistic value of 0.8 for the reflectance at the rear side, the planar and 1000 nm pitch nanohole array reach 33.0\% and 58.5\% transmittance respectively.

The case of a diffuse, i.e. Lambertian type reflector is presented in figure \ref{fig:MultiPassTransmission}(b). The extrema of the curves are similar to the those in (a), with the notable except that planar now increases to 100\% transmittance for unity reflectance. The angular dependence of the transmittance is signicantly altered. The dependence for planar and each of the nanohole arrays is nearly indistinguishable. This significant departure from the angular dependence in (a) is due to the diffuse reflector's ability to redistribute the angular distribution of light trapped inside the AlN layer. Light which is intially propagating at a high angle to the interface normal which would be trapped in the planar case or weakly transmitted by the nanohole array is now efficiently scattered into low angles which have a high transmittance in both the planar and nanostructured case, as can be seen in figure \ref{fig:SinglePassTransmission}. This process can more efficiently extract the trapped light, leading to higher values of transmittance for a given value of reflectance compared to the specular reflector. If a diffuse rear reflector is available then nanostructuring the AlN/{\sapphire} interface would have no optical benefit compared to planar.

\subsection{Total Light Extraction}
\begin{figure*}[htbp]
	\centering
	\includegraphics[width=\hdoublecolw]{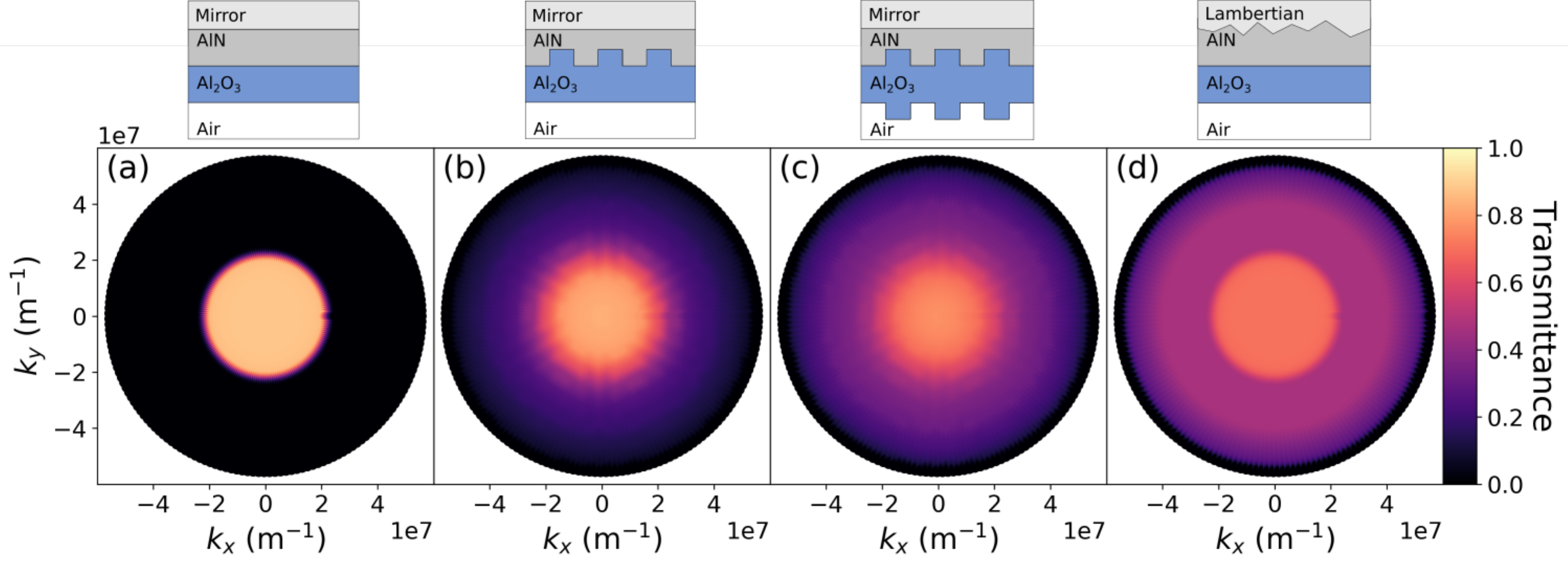}
	\caption{Transmittance into air for in emission in the AlN layer as a function of {\ktrans}. The inset in each case shows a sketch of the device  structure. (a) Mirror, planar AlN/{\sapphire} interface, planar {\sapphire}/air interface; (b) mirror, nanostructured AlN/{\sapphire} interface, planar {\sapphire}/air interface; (c) mirror, nanostructured AlN/{\sapphire} interface, nanostructured {\sapphire}/air interface; (d) Lambertian scatterer, planar AlN/{\sapphire} interface, planar {\sapphire}/air interface.}
	\label{fig:FullStackAngularTransmission}
\end{figure*}

Having laid the groundwork in the previous two sections we are now in the position to analyse the LEE of the NPSS for extraction into air. Until this point we have considered extraction of light from the AlN emission layer into the {\sapphire} substrate. Clearly for a fully functioning device the light should be extracted from the substrate into air. The interface between AlN and {\sapphire} needs to be carefully considered, since the morphology there has a large impact not only on the optical properties but also on the material quality of the AlN that is grown on top. On the other hand, the lower surface of the {\sapphire} substrate can be modified for maximum optical benefit since it is essentially decoupled from the rest of the device. A multitude of geometries useful for light extraction could be used at the lower interface. In the current work we present the same nanohole structure used at the AlN/{\sapphire} interface at the {\sapphire}/air interface as a light extraction concept. This has the advantage that both sides of the {\sapphire} substrate need be to patterned in the same way

Figure \ref{fig:FullStackAngularTransmission} shows the transmittance into air as a function of the transversal components of $\vec{k}$ for emission inside of the AlN layer. The transmittance value is the averaged over TE and TM polarization and over upwards and and downwards emission. A mirror reflector with $R = 0.8$ has been taken at the rear of the device. This could be in the form of highly reflective Ni/Mg \cite{Khan2019} or Rh electrodes \cite{Takano2017}. Four different cases are considered. Firstly if all interfaces in the device are planar, secondly if only the upper AlN/{\sapphire} interface is structured, thirdly if both the upper AlN/{\sapphire} and lower {\sapphire}/air interface are structured, and finally if the mirror is exchange for a diffuse reflector and all other layers are planar. The integrated values of transmittance are presented in table \ref{tab:FullStackTable}.

\begin{table}[htbp]
	\centering
	\begin{tabular}{c|c|c|c}
		Reflector & \begin{tabular}{@{}c@{}} AlN/{\sapphire} \\ Interface\end{tabular} 
		&  \begin{tabular}{@{}c@{}} {\sapphire}/Air \\ Interface\end{tabular} & Transmittance \\
		\hline
		Specular & Planar & Planar & 7.8\% \\ 
		Specular & Nanoholes & Planar & 19.0\% \\
		Specular & Nanoholes & Nanholes & 25.0\% \\
		Diffuse & Planar & Planar  & 34.3\% \\
	\end{tabular}
	\caption{The integrated transmittance for the data presented in figure \ref{fig:FullStackAngularTransmission}}
	\label{tab:FullStackTable}
\end{table}

The fully planar device shows a high transmittance at low emission angles, mainly limited by the 0.8 reflectance of the rear reflector. The transmittance falls rapidly to zero at $\vec{k}$ value corresponding to the critical angle of 25\si{\degree}, which has been reduced from 51\si{\degree} due to the {\sapphire}/air interface. Due to higher emission angles containing more energy, the total transmittance for the fully planar device is only 7.8\%. The situation changes when the upper interface is structured with the 1000 nm pitch nanohole array. The transmittance distribution is smoothed out covering the whole range of $\vec{k}$ values. The distribution is smoother than the curve in figure \ref{fig:SinglePassTransmission} due to multiple interactions with the structured interface. The integrated transmittance for the upper nanohole array plus lower planar interface is 19.0\%, more than double the fully planar case. By adding a further nanohole array at the lower interface (c) the transmittance distribution is further smoothed with only a minimal drop of transmittance at the critical angle. The transmittance at small $\vec{k}$ is lower than the previous two cases but the high $\vec{k}$ transmittance is the much higher. This leads to an integrated transmittance of 25.0\% for the device with two nanohole array interfaces. The diffuse scatter (d) achieves 34.0\% transmission even with all other layers being planar. This highlights the fact that a single structured interface can extract light efficiently if it can scattering light from high to low angles, therefore reducing device complexity.

All these values are limited by the reflectance of the rear interface, i.e. by optical losses in the system. With higher values of reflectance for the rear reflector, the total transmittance is expected to grow nonlinearly towards 100.0\% as in figure \ref{fig:MultiPassTransmission}. This highlights the need for reduced optical losses in DUV-LEDs to make use of light scattering provided by nanopatterned sapphire substrates.

\section{Conclusion}

We presented an analysis of light extraction from deep-UV light emitting diodes based on AlN on nanopatterned sapphire substrates. To perform this analysis we developed an optical model combining rigorously computed scattering matrices describing interfaces with a multilayer solver. Brillouin zone sampling in Bloch families allowed us to reduce the amount of computational effort needed to calculate the scattering matrix considerably. 

We first compared the transmittance from AlN to sapphire for three kinds of nanostructure with pitches ranging from 265\,nm to 1000\,nm and a planar interface. We found that in the case of a fully absorbing rear contact in the device, hence only a single light interaction with the AlN/sapphire interface, transmittance only marginally increases by nanopatterning. However, by eliminating the parasitic absorption and adding a mirror to the rear side of the device we find significantly increased transmittance into the {\sapphire} substrate by nanopatterning. We found the highest transmittance for the largest pitches investigated. 

As an example, we calculated the total light extraction from the whole device assuming a transparent p-side and a rear-side mirror with 0.8 reflectance. For the all planar device the total light extraction was 7.8\%, rising to 19.0\% when the AlN/sapphire interface was nanostructured and again increasing to 25.0\% when both the AlN/sapphire and sapphire/air interfaces were nanostructured with a 1000\,nm periodic nanohole array. We further introduced a novel method using a diffuse ('Lambertian') mirror at the rear-side of the device, enabling light extraction efficiency of 34.2\% even if all other interfaces are planar.  This supports the idea that only a single scattering layer is needed for high light extraction, if the scattering pathway from high angles to lower angles is efficient enough. This could allow for more low cost designs, since fewer nanostructuring steps need to be taken. Crucially, this consideration depends on a low optical loss environment inside the LED thin film. This highlights the need for continued research into transparent p-sides and highly reflective p-contacts for DUV-LED devices. Until these can be obtained, the power of nanostructuring to increase the light extraction efficiency will be largely left untapped.

The optical design guidelines presented in this study can help to accelerate the development of highly efficient DUV-LEDs, but the model is also readily applicable to other multi-layer opto-electronic nanostructured devices such as photovoltaics or photodetectors. 

\section*{Funding Information}
Partially funded by the Deutsche Forschungsgemeinschaft (DFG, German Research Foundation) under Germany´s Excellence Strategy – The Berlin Mathematics Research Center MATH+ (EXC-2046/1, project ID: 390685689, AA4-6), and by the German Federal Ministry of Education and Research (BMBF) within the Advanced UV for Life project consortium and by the Helmholtz Innovation Lab HySPRINT, which is financially supported by the Helmholtz Association. The simulations were done at the Berlin Joint Lab for Optical Simulations for Energy Research (BerOSE) and at the Helmholtz Excellence Cluster SOLARMATH, a strategic collaboration of Helmholtz-Zentrum Berlin and MATH+.

\section*{Acknowledgements}
We thank Prof.\ Markus Weyers for critically revising the manuscript and Dr. Carlo Barth for initial setup of the optical simulations. 

\section*{Disclosures}
The authors declare that there are no conflicts of interest related to this article.

\vspace{\baselineskip} See Supplement 1 for supporting content.
\smallskip
\newpage
\section{Supporting Information}
\label{Supp:sec}

\subsection{Scattering Matrix Definition}
\label{Supp:subsec:smatrixdef}

\begin{figure}
	\centering
	\includegraphics[height=40mm]{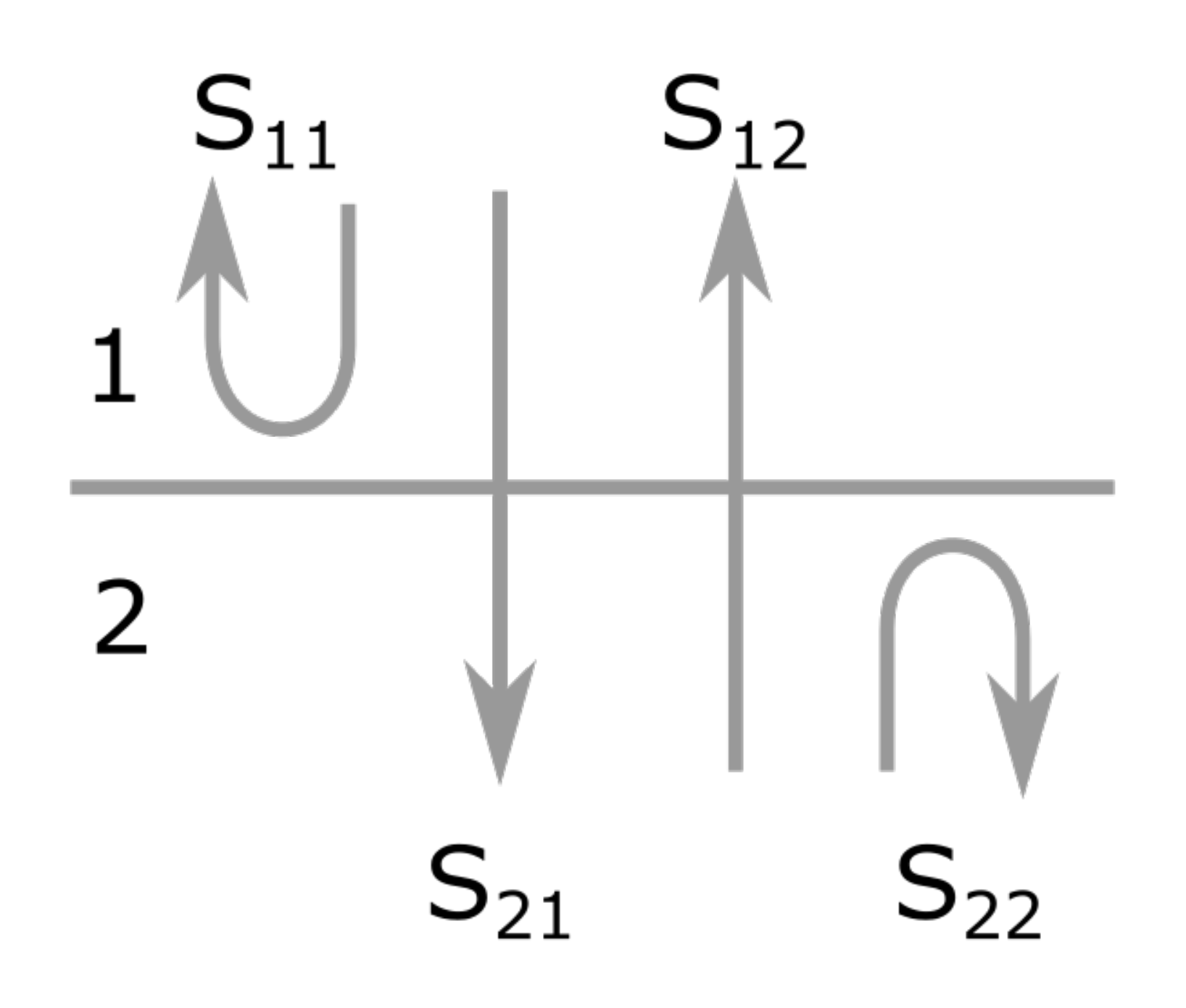}
	\caption{Schematic of the four components of a scattering matrix. Typically each component is itself a matrix describing the coupling of modes in layers 1 and 2 at the interface.}
	\label{fig:smatrix_1D}
\end{figure}

The scattering matrix describes how light is reflected and transmitted at a material interface. Figure \ref{fig:smatrix_1D} schematically shows the four cases to be considered. With no loss of generality we define the interface to lie in the $x-y$ plane and consider light propagating with some component of the $k$ vector $\vec{k}$ along the optical axis $z$. Propagation of light purely in the $x-y$ plane describes evanescent modes. For extremely large values of $|\vec{k}|$ the propagation approaches a singular, localised field which does not propagate. Since we assume the source is sufficiently far away from the interface described using scattering matrices, these modes may be neglected. We assume layers 1 and 2 are above and below the interface, respectively. The {\Soneone} sub-matrix describes the reflected light for incidence from layer 1. Likewise {\Stwoone} describes transmission from layer 1 to 2, {\Sonetwo} describes transmission from layer 2 to 1 and {\Stwotwo} describes reflection for light incident from layer 2. Generally we can describe the components of the scattering matrix as being transfer conditions between the modes in the upper and lower layers.

The shapes of the {\Snm} sub-matrices are determined by the choice of plane waves propagating in each homogeneous half space. As an example, if only a single mode is considered then the {\Snm} conditions are scalars describing the reflection and transmission of the single mode. Consequently the entire scattering matrix has size $2\times2$. Furthermore if the interface between layer one and two is planar then these coupling conditions are given by the well known Fresnel equations for reflection and transmission. 

More generally, two linearly independent polarisations as well as multiple incident angles are used. This provides a basis of modes able to describe arbitrary light distributions. Given $N$ and $M$ different angles in layers 1 and 2, respectively, the sub-matrices {\Soneone}, {\Sonetwo}, {\Stwoone} and {\Stwotwo} become $2N\times2N$, $2M\times2N$, $2N\times2M$ and $2M\times2M$ sized, respectively. The entire scattering matrix will then have size $2(N+M)\times2(N+M)$.

\subsection{Obtaining Coupling Coefficients}
Here we will describe the method used to obtain the coupling coefficients which constitute the individual entries in the scattering matrix using FEM. Firstly, the scattering problem is solved for a system containing the nanostructured interface between two homogeneous half spaces. Periodic boundary conditions are used in the $x-y$ plane while transparent boundary conditions are used in the $z$ direction. As source term a plane wave illumination is used with a given angle of incidence and polarization. In order to obtain the amplitudes of the outgoing plane waves, the Fourier transform of the electric field propagating outwards in the $+z$ and $-z$ directions can be taken on the computational domain boundary. By normalizing the outgoing amplitudes with the incoming amplitude, the coupling coefficient between incident and outgoing plane waves is obtained. This process is then repeated for different values of the incident angle and polarization (for both incidence from above and below the interface) until the coupling coefficients between all combinations of incident and outgoing plane waves have been determined.

All simulations were performed with vacuum wavelength of 265\,nm. The refractive indices of AlN and {\sapphire} used for simulations were 2.35 and 1.83, respectively. These values were taken from literature \cite{Pastrnak1966,Dodge1986}. Optical losses in AlN and {\sapphire} were assumed to be zero. Convergence tests were performed to ensure that simulations were accurate to at least $10^{-2}$. The mesh was set to have maximum side lengths equal to 0.25 of the wavelength of light in each given material. The finite element degree was set adaptively in each element to ensure the aforementioned accuracy \cite{Pomplun2007}. Typically, values of 3-4 for the finite element degree were sufficient to reach the required accuracy.

\begin{figure}
	\centering
	\includegraphics[width=\hcolw]{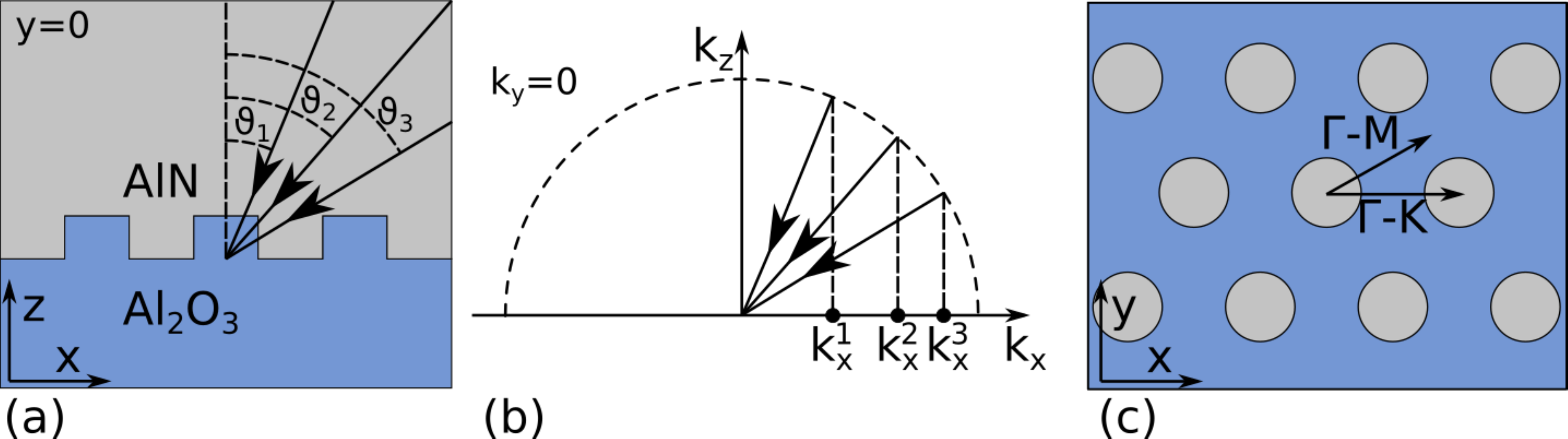}
	\caption{(a) Schematic of a nanostructured AlN/{\sapphire} interface in the $x-z$ plane. The different values of incident polar angle $\theta$ of light are show. (b) The reciprocal space representation of (a), the semicircle represents the cutoff wavenumber $k_{\mathrm{c}}$ for AlN, the three different incident $k$ vectors are equivalent to the incident angles in (a). Only the $k_{x}-k_{z}$ plane is shown due to $k_{y}$ being zero for the incident $k$ vectors shown, in general the semicircle will form a hemisphere in the $k_{z} > 0$ half of reciprocal space. (c) Schematic of the nanostructure shown in (a) now seen from above. The high symmetry directions of the triangular lattice are indicated.}
	\label{fig:Angles}
\end{figure}

\subsection{Choice of Modes}
Due to the assumption of homogeneous layers between the interfaces, the propagation of light in the layers will be described via plane waves. The combination of angles and polarizations of these plane waves form a basis for describing arbitrary angular distributions of light. For the polarization state of the light, two linearly independent polarizations are sufficient. On the other hand, the angular distribution of the light contains a continuum of modes. In practice a finite number of modes is typically sufficient to reconstruct the angular distribution with acceptable accuracy. In the following we will demonstrate how carefully choosing the modes can lead to an enormous decrease in the time required for computing the scattering matrices.

In order to describe the choice of modes, it is convenient to change from describing the different modes as angles of propagation and instead use the k vector components in the $x-y$ plane, {\ktrans}$ = (k_{x},k_{y})$. The correspondence between angles and {\ktrans} components is shown in figure \ref{fig:Angles}(a-b). Considering light incident to an interface in the $x-z$ plane, the azimuthal angle $\phi$ (shown in figure 1 in main text) will be zero for all incident angles. The polar angle $\theta$ in (a) is shown for three different values corresponding to three values of $k_{x}$ in (b). In general the values of {\ktrans} will be given by,

\begin{align}
k_{x} &= nk_{0}\cos\phi \sin\theta, \nonumber\\
k_{y} &= nk_{0}\sin\phi \sin\theta.
\end{align}
where $n$ is the refractive index of the layer and $k_{0}$ is the wavenumber of the plane wave in vacuum.

\begin{figure}
	\centering
	\includegraphics[width=0.8\hcolw]{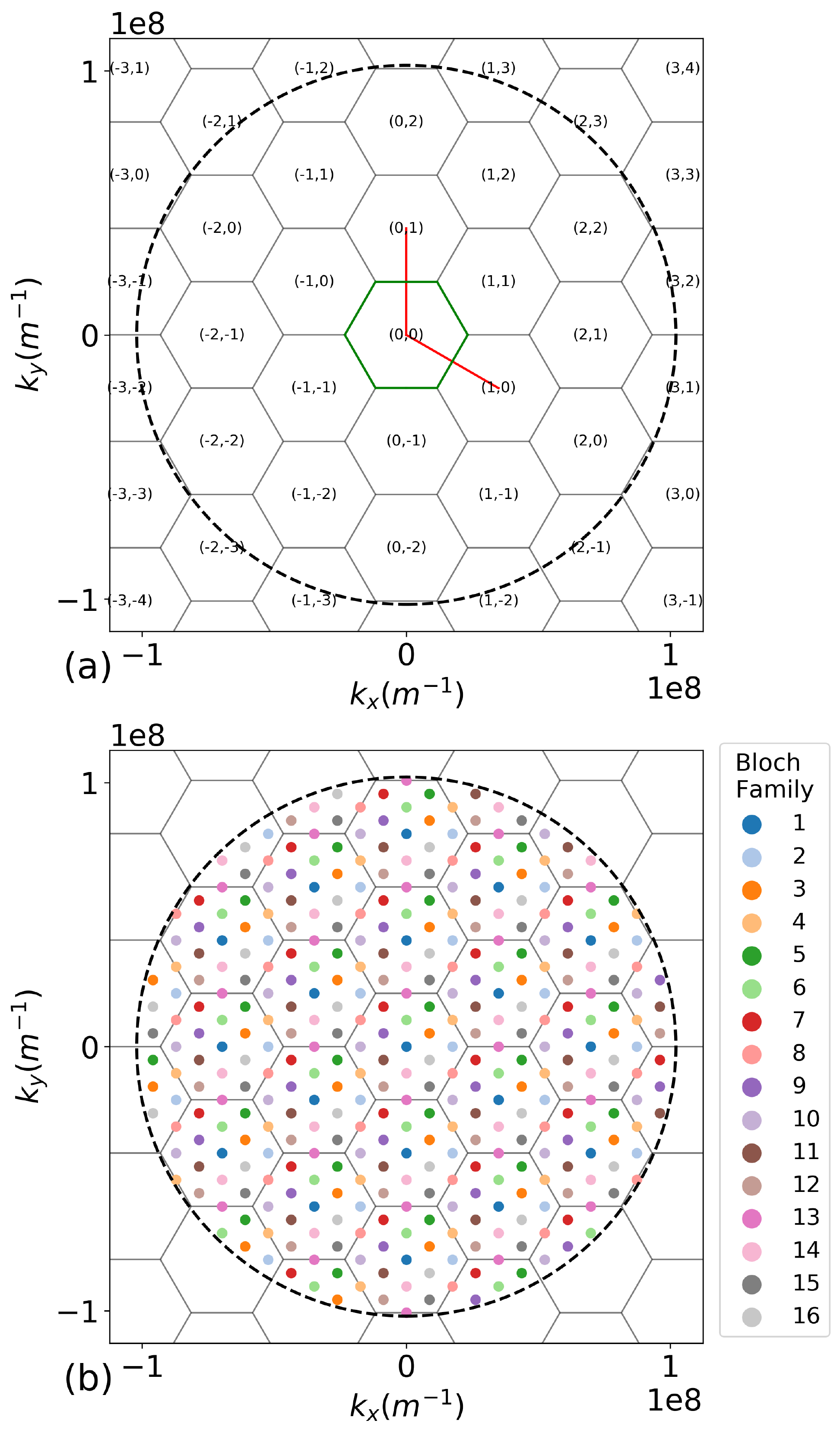}
	\caption{The circular numerical aperture in reciprocal space is shown in both (a) and (b) for sapphire at 265 nm vacuum wavelength. In (a) the reciprocal lattice of a hexagonal periodic structure with 180 nm pitch has been overlaid, as well as the two reciprocal lattice vectors (red lines). The first Brillouin zone is highlighted in green. In (b) the sampling of points inside $k_{\mathrm{c}}$ have been color coded with the associated Bloch family. The pattern of 16 points is periodically repeated in each unit cell of the reciprocal lattice.}
	\label{fig:KSamplingBlochFamilies}    
\end{figure}

The values of the components {\ktrans} are restricted by the maximum value of {\kvector} inside the layer,
\begin{equation}
\|\vec{k_{\perp}}\| < \|\vec{k}\|.
\end{equation}
This forms a circle in reciprocal space inside which the values of {\ktrans} need to be sampled. Outside of this circle lie the evanescent modes which are needed for a general description of the interface. Under the assumption that light sources are sufficiently far from the interface as to not excite evanescent modes, we can neglect the evanescent modes. Since {\kvector} is dependent on the refractive index of the layer, the radius of the circle will be different in each layer. We will refer to the radius of the circle in reciprocal as the cutoff wave wavenumber $k_{\mathrm{c}}$. Figure \ref{fig:KSamplingBlochFamilies}(a) and (b) show examples of the $k_{\mathrm{c}}$ circle in reciprocal space.

If the interface is periodically structured, such as the example in figure \ref{fig:Angles}(c), then the sampling of points can be chosen in such a way as to reduce the computational costs of the calculation. 

Figure \ref{fig:KSamplingBlochFamilies}(a) shows the reciprocal lattice for a hexagonal periodic structure overlaid onto the $k_{\mathrm{c}}$ circle in reciprocal space for sapphire at 265 nm wavelength. The reciprocal lattice is formed from a tessellation of hexagonal Brillouin zones, with the first Brillouin zone being highlighted in the center, as well as the two reciprocal lattice vectors used to generate the lattice.

Any point lying outside of the first Brillouin zone may be reached via taking a point within the first Brillouin zone and translating it by adding appropriate multiples of the two reciprocal lattice vectors. Consequently, for each unique {\ktrans} inside the first Brillouin zone, a set of {\ktrans} inside $k_{\mathrm{c}}$ exists which are all separated by an integer multiple of the reciprocal lattice vectors. We will refer to this set of points as a Bloch family. Figure \ref{fig:KSamplingBlochFamilies}(b) shows an example of 16 such Bloch families.

For incident plane waves with {\ktrans} in a given Bloch family, the finite element system matrix used to compute the solution will be the same. This means that the same system matrix can be inverted once and the inversion used to get the solution for all plane waves \cite{Burger2013}. Inversion of the system matrix is the most computationally expensive part of a FEM. Therefore, reducing the frequency with which this must be performed can have a large impact on the computation time. Writing the total computational time as
\begin{equation}
t_{\mathrm{total}} = t_{\mathrm{inv}} + t_{\mathrm{0}},
\end{equation}
where $t_{\mathrm{inv}}$ is the time for the system matrix inversion and $t_{\mathrm{0}}$ accounts for all other operations performed, in the limit $t_{\mathrm{inv}} >> t_{\mathrm{0}}$ the unitless speed up in $t_{\mathrm{total}}$, $\Delta_{\mathrm{total}}$, will be given by the speed up in $t_{\mathrm{inv}}$, $\Delta_{\mathrm{inv}}$. This will be determined by the ratio of the area of the $k_{\mathrm{c}}$ circle to that of the first Brillouin zone. For a hexagonal periodic lattice with pitch $P$, $\Delta_{\mathrm{inv}}$ will be given by,
\begin{equation}
\Delta_{\mathrm{inv}} = \frac{2\pi*n^{2}P^{2}}{\sqrt{3}\lambda^{2}},
\end{equation}
where $n$ is the refractive index of the material at the interface and $\lambda$ is the vacuum wavelength of the incident light. Since $\Delta_{\mathrm{inv}}$ is dependent on the refractive index, the values of $\Delta_{\mathrm{inv}}$ for light incident from the upper and lower materials should be averaged to estimate the speed up of the computation of a scattering matrix at a material interface. 

Considering only plane waves incident from the AlN side of an AlN/{\sapphire} interface consisting of a 1000 nm pitch nanohole array at 265 nm wavelength, the upper limit on  $\Delta_{\mathrm{total}}$ is 285. The value obtained for realistic calculations was 88 which is still a significant speed up, but highlights the fact that $t_{0}$ cannot be neglected in this case.

\subsection{Brillouin Zone Sampling}
The sampling of points inside the first Brillouin zone needs to be defined in order to obtain the Bloch families. There is a rich literature on optimal sampling of the Brillouin zone, which have commonly been used for quantum mechanical calculations involving crystal structures. Given a sampling of the irreducible Brillouin zone, we obtain the sampling over the whole Brillouin zone via appropriate symmetry operations. In this work we chose a Monkhurst-Pack grid for sampling the irreducible Brillouin zone \cite{Monkhorst1976}.
This spaces points equidistantly along directions parallel to the reciprocal lattice vectors. 

\newpage
\iftoggle{usebiber}
{
	\printbibliography
}
{
	\bibliography{LEDpaper2019}

@article{Kashima2017,
	doi = {10.7567/apex.11.012101},
	url = {https://doi.org/10.7567\%2Fapex.11.012101},
	year = 2017,
	month = {nov},
	publisher = {Japan Society of Applied Physics},
	volume = {11},
	number = {1},
	pages = {012101},
	author = {Yukio Kashima and Noritoshi Maeda and Eriko Matsuura and Masafumi Jo and Takeshi Iwai and Toshiro Morita and Mitsunori Kokubo and Takaharu Tashiro and Ryuichiro Kamimura and Yamato Osada and Hideki Takagi and Hideki Hirayama},
	title = {High external quantum efficiency (10{\%}) {AlGaN}-based deep-ultraviolet light-emitting diodes achieved by using highly reflective photonic crystal on p-{AlGaN} contact layer},
	journal = {Applied Physics Express}
}

@inproceedings{Burger2005,
	author = {Sven Burger and Roderick Köhle and Lin Zschiedrich and Weimin Gao and Frank Schmidt and Reinhard März and Christoph Nölscher},
	title = {{Benchmark of FEM, waveguide, and FDTD algorithms for rigorous mask simulation}},
	volume = {5992},
	booktitle = {25th Annual BACUS Symposium on Photomask Technology},
	editor = {J. Tracy Weed and Patrick M. Martin},
	organization = {International Society for Optics and Photonics},
	publisher = {SPIE},
	pages = {378 -- 389},
	year = {2005},
	doi = {10.1117/12.631696},
	URL = {https://doi.org/10.1117/12.631696}
}

@phdthesis{Hammerschmidt2016,
  author      = {Martin Hammerschmidt},
  title       = {Optical simulation of complex nanostructured solar cells with a reduced basis method},
  school = {Free University Berlin},
  pages       = {XX, 169},
  url       = {http://www.diss.fu-berlin.de/diss/receive/FUDISS_thesis_000000102429},
  year        = {2016},
}

@article{Hirayama2014,
	author = {Hideki Hirayama and Noritoshi Maeda and Sachie Fujikawa and Shiro Toyoda and Norihiko Kamata},
	doi = {10.7567/jjap.53.100209},
	journal = {Japanese Journal of Applied Physics},
	month = {sep},
	number = {10},
	pages = {100209},
	publisher = {Japan Society of Applied Physics},
	title = {Recent progress and future prospects of {AlGaN}-based high-efficiency deep-ultraviolet light-emitting diodes},
	url = {https://doi.org/10.7567\%2Fjjap.53.100209},
	volume = {53},
	year = 2014
}

@article{Cho2014,
author = {Doo-Hee Cho and Jin-Wook Shin and Chul Woong Joo and Jonghee Lee and Seung Koo Park and Jaehyun Moon and Nam Sung Cho and Hye Yong Chu and Jeong-Ik Lee},
journal = {Opt. Express},
number = {S6},
pages = {A1507--A1518},
publisher = {OSA},
title = {Light diffusing effects of nano and micro-structures on {OLED} with microcavity},
volume = {22},
month = {Oct},
year = {2014},
url = {http://www.opticsexpress.org/abstract.cfm?URI=oe-22-106-A1507},
doi = {10.1364/OE.22.0A1507},
}

@article{Matioli2010,
	title = {Impact of photonic crystals on {LED} light extraction efficiency: approaches and limits to vertical structure designs},
	volume = {43},
	issn = {0022-3727, 1361-6463},
	shorttitle = {Impact of photonic crystals on {LED} light extraction efficiency},
	url = {http://stacks.iop.org/0022-3727/43/i=35/a=354005?key=crossref.a4996fa06717e070f3e9b4cdf40d57cc},
	doi = {10.1088/0022-3727/43/35/354005},
	language = {en},
	number = {35},
	urldate = {2019-05-20},
	journal = {Journal of Physics D: Applied Physics},
	author = {Matioli, Elison and Weisbuch, Claude},
	month = sep,
	year = {2010},
	pages = {354005}
}

@article{Kim2005,
	title = {Enhanced light extraction from {GaN}-based light-emitting diodes with holographically generated two-dimensional photonic crystal patterns},
	volume = {87},
	issn = {0003-6951, 1077-3118},
	url = {http://aip.scitation.org/doi/10.1063/1.2132073},
	doi = {10.1063/1.2132073},
	language = {en},
	number = {20},
	urldate = {2019-05-20},
	journal = {Applied Physics Letters},
	author = {Kim, Dong-Ho and Cho, Chi-O and Roh, Yeong-Geun and Jeon, Heonsu and Park, Yoon Soo and Cho, Jaehee and Im, Jin Seo and Sone, Cheolsoo and Park, Yongjo and Choi, Won Jun and Park, Q-Han},
	month = nov,
	year = {2005},
	pages = {203508}
}

@Article{Gutierrez2018,
AUTHOR = {Gutiérrez, Yael and Alcaraz de la Osa, Rodrigo and Ortiz, Dolores and Saiz, José María and González, Francisco and Moreno, Fernando},
TITLE = {Plasmonics in the Ultraviolet with Aluminum, Gallium, Magnesium and Rhodium},
JOURNAL = {Applied Sciences},
VOLUME = {8},
YEAR = {2018},
NUMBER = {1},
ARTICLE-NUMBER = {64},
URL = {https://www.mdpi.com/2076-3417/8/1/64},
ISSN = {2076-3417},
DOI = {10.3390/app8010064}
}

@article{Akcay2002,
author = {Ceyhun Akcay and Pascale Parrein and Jannick P. Rolland},
journal = {Appl. Opt.},
number = {25},
pages = {5256--5262},
publisher = {OSA},
title = {Estimation of longitudinal resolution in optical coherence imaging},
volume = {41},
month = {Sep},
year = {2002},
url = {http://ao.osa.org/abstract.cfm?URI=ao-41-25-5256},
doi = {10.1364/AO.41.005256},
}

@Article{Nagasawa2018,
AUTHOR = {Nagasawa, Yosuke and Hirano, Akira},
TITLE = {A Review of {AlGaN}-Based Deep-Ultraviolet Light-Emitting Diodes on Sapphire},
JOURNAL = {Applied Sciences},
VOLUME = {8},
YEAR = {2018},
NUMBER = {8},
ARTICLE-NUMBER = {1264},
URL = {https://www.mdpi.com/2076-3417/8/8/1264},
ISSN = {2076-3417},
DOI = {10.3390/app8081264}
}

@article{Ryu2013,
	doi = {10.7567/apex.6.062101},
	url = {https://doi.org/10.7567\%2Fapex.6.062101},
	year = 2013,
	month = {jun},
	publisher = {Japan Society of Applied Physics},
	volume = {6},
	number = {6},
	pages = {062101},
	author = {Han-Youl Ryu and Il-Gyun Choi and Hyo-Sik Choi and Jong-In Shim},
	title = {Investigation of Light Extraction Efficiency in {AlGaN} Deep-Ultraviolet Light-Emitting Diodes},
	journal = {Applied Physics Express}
}

@article{Pernot2010,
	doi = {10.1143/apex.3.061004},
	url = {https://doi.org/10.1143\%2Fapex.3.061004},
	year = 2010,
	month = {jun},
	publisher = {{IOP} Publishing},
	volume = {3},
	number = {6},
	pages = {061004},
	author = {Cyril Pernot and Myunghee Kim and Shinya Fukahori and Tetsuhiko Inazu and Takehiko Fujita and Yosuke Nagasawa and Akira Hirano and Masamichi Ippommatsu and Motoaki Iwaya and Satoshi Kamiyama and Isamu Akasaki and Hiroshi Amano},
	title = {Improved Efficiency of 255{\textendash}280 nm {AlGaN}-Based Light-Emitting Diodes},
	journal = {Applied Physics Express}
}

@Article{Khan2019,
author ="Khan, M. Ajmal and Maeda, Noritoshi and Jo, Masafumi and Akamatsu, Yuki and Tanabe, Ryohei and Yamada, Yoichi and Hirayama, Hideki",
title  ="13 {mW} operation of a 295–310 nm {AlGaN} {UV-B} {LED} with a p-{AlGaN} transparent contact layer for real world applications",
journal  ="J. Mater. Chem. C",
year  ="2019",
volume  ="7",
issue  ="1",
pages  ="143-152",
publisher  ="The Royal Society of Chemistry",
doi  ="10.1039/C8TC03825B",
url  ="http://dx.doi.org/10.1039/C8TC03825B"}

@article{Oder2003,
	title = {{III}-nitride photonic crystals},
	volume = {83},
	issn = {0003-6951, 1077-3118},
	url = {http://aip.scitation.org/doi/10.1063/1.1600839},
	doi = {10.1063/1.1600839},
	language = {en},
	number = {6},
	urldate = {2019-05-20},
	journal = {Applied Physics Letters},
	author = {Oder, T. N. and Shakya, J. and Lin, J. Y. and Jiang, H. X.},
	month = aug,
	year = {2003},
	pages = {1231--1233}
}

@article{Monkhorst1976,
	title = {Special points for {Brillouin}-zone integrations},
	volume = {13},
	issn = {0556-2805},
	url = {https://link.aps.org/doi/10.1103/PhysRevB.13.5188},
	doi = {10.1103/PhysRevB.13.5188},
	language = {en},
	number = {12},
	urldate = {2019-05-20},
	journal = {Physical Review B},
	author = {Monkhorst, Hendrik J. and Pack, James D.},
	month = jun,
	year = {1976},
	pages = {5188--5192}
}

@article{Becker2015,
	title = {5 × 5 cm$^{2}$ silicon photonic crystal slabs on glass and plastic foil exhibiting broadband absorption and high-intensity near-fields},
	volume = {4},
	issn = {2045-2322},
	url = {http://www.nature.com/articles/srep05886},
	doi = {10.1038/srep05886},
	language = {en},
	number = {1},
	urldate = {2019-05-20},
	journal = {Scientific Reports},
	author = {Becker, C. and Wyss, P. and Eisenhauer, D. and Probst, J. and Preidel, V. and Hammerschmidt, M. and Burger, S.},
	month = may,
	year = {2015},
	pages = {5886}
}

@article{Hollander2018,
	title = {{3D} printed {UV} light cured polydimethylsiloxane devices for drug delivery},
	volume = {544},
	issn = {0378-5173},
	url = {http://www.sciencedirect.com/science/article/pii/S0378517317310712},
	doi = {https://doi.org/10.1016/j.ijpharm.2017.11.016},
	number = {2},
	journal = {International Journal of Pharmaceutics},
	author = {Holländer, Jenny and Hakala, Risto and Suominen, Jaakko and Moritz, Niko and Yliruusi, Jouko and Sandler, Niklas},
	year = {2018},
	pages = {433 -- 442}
}

@article{Kim2011,
	doi = {10.1143/apex.4.092102},
	url = {https://doi.org/10.1143\%2Fapex.4.092102},
	year = 2011,
	month = {aug},
	publisher = {{IOP} Publishing},
	volume = {4},
	number = {9},
	pages = {092102},
	author = {Myunghee Kim and Takehiko Fujita and Shinya Fukahori and Tetsuhiko Inazu and Cyril Pernot and Yosuke Nagasawa and Akira Hirano and Masamichi Ippommatsu and Motoaki Iwaya and Tetsuya Takeuchi and Satoshi Kamiyama and Masahito Yamaguchi and Yoshio Honda and Hiroshi Amano and Isamu Akasaki},
	title = {{AlGaN}-Based Deep Ultraviolet Light-Emitting Diodes Fabricated on Patterned Sapphire Substrates},
	journal = {Applied Physics Express}
}

@article{Riesen2018,
	title = {Towards rewritable multilevel optical data storage in single nanocrystals},
	volume = {26},
	issn = {1094-4087},
	url = {https://www.osapublishing.org/abstract.cfm?URI=oe-26-9-12266},
	doi = {10.1364/OE.26.012266},
	language = {en},
	number = {9},
	urldate = {2019-05-20},
	journal = {Optics Express},
	author = {Riesen, Nicolas and Pan, Xuanzhao and Badek, Kate and Ruan, Yinlan and Monro, Tanya M. and Zhao, Jiangbo and Ebendorff-Heidepriem, Heike and Riesen, Hans},
	month = apr,
	year = {2018},
	pages = {12266}
}

@article{Jones2014,
	title = {{UV} {Light} {Inactivation} of {Human} and {Plant} {Pathogens} in {Unfiltered} {Surface} {Irrigation} {Water}},
	volume = {80},
	issn = {0099-2240, 1098-5336},
	url = {http://aem.asm.org/lookup/doi/10.1128/AEM.02964-13},
	doi = {10.1128/AEM.02964-13},
	language = {en},
	number = {3},
	urldate = {2019-05-20},
	journal = {Applied and Environmental Microbiology},
	author = {Jones, Lisa A. and Worobo, Randy W. and Smart, Christine D.},
	month = feb,
	year = {2014},
	pages = {849--854}
}

@article{Yin2013,
	title = {Light based anti-infectives: ultraviolet {C} irradiation, photodynamic therapy, blue light, and beyond},
	volume = {13},
	issn = {1471-4892},
	url = {http://www.sciencedirect.com/science/article/pii/S1471489213001550},
	doi = {https://doi.org/10.1016/j.coph.2013.08.009},
	number = {5},
	journal = {Current Opinion in Pharmacology},
	author = {Yin, Rui and Dai, Tianhong and Avci, Pinar and Jorge, Ana Elisa Serafim and Melo, Wanessa CMA de and Vecchio, Daniela and Huang, Ying-Ying and Gupta, Asheesh and Hamblin, Michael R.},
	year = {2013},
	pages = {731 -- 762}
}

@INPROCEEDINGS{Goetzberger1981,
   author = {{Goetzberger}, A.},
    title = "{Optical confinement in thin Si-solar cells by diffuse back reflectors}",
booktitle = {15th Photovoltaic Specialists Conference},
     year = 1981,
    pages = {867-870},
   adsurl = {https://ui.adsabs.harvard.edu/abs/1981pvsp.conf..867G}
}

@article{Yablonovitch1982,
author = {Eli Yablonovitch},
journal = {J. Opt. Soc. Am.},
number = {7},
pages = {899--907},
publisher = {OSA},
title = {Statistical ray optics},
volume = {72},
month = {Jul},
year = {1982},
url = {http://www.osapublishing.org/abstract.cfm?URI=josa-72-7-899},
doi = {10.1364/JOSA.72.000899},
}

@article{Saxena2009,
	title = {A review on the light extraction techniques in organic electroluminescent devices},
	volume = {32},
	issn = {09253467},
	url = {https://linkinghub.elsevier.com/retrieve/pii/S0925346709002316},
	doi = {10.1016/j.optmat.2009.07.014},
	language = {en},
	number = {1},
	urldate = {2019-05-20},
	journal = {Optical Materials},
	author = {Saxena, Kanchan and Jain, V.K. and Mehta, Dalip Singh},
	month = nov,
	year = {2009},
	pages = {221--233}
}

@article{Gessmann2004,
	title = {High-efficiency {AlGaInP} light-emitting diodes for solid-state lighting applications},
	volume = {95},
	issn = {0021-8979, 1089-7550},
	url = {http://aip.scitation.org/doi/10.1063/1.1643786},
	doi = {10.1063/1.1643786},
	language = {en},
	number = {5},
	urldate = {2019-05-20},
	journal = {Journal of Applied Physics},
	author = {Gessmann, Th. and Schubert, E. F.},
	month = mar,
	year = {2004},
	pages = {2203--2216}
}

@article{Windisch1999,
	title = {Light-emitting diodes with 31\% external quantum efficiency by outcoupling of lateral waveguide modes},
	volume = {74},
	issn = {0003-6951, 1077-3118},
	url = {http://aip.scitation.org/doi/10.1063/1.123817},
	doi = {10.1063/1.123817},
	language = {en},
	number = {16},
	urldate = {2019-05-20},
	journal = {Applied Physics Letters},
	author = {Windisch, R. and Heremans, P. and Knobloch, A. and Kiesel, P. and Döhler, G. H. and Dutta, B. and Borghs, G.},
	month = apr,
	year = {1999},
	pages = {2256--2258}
}

@article{Fujii2004,
	title = {Increase in the extraction efficiency of {GaN}-based light-emitting diodes via surface roughening},
	volume = {84},
	issn = {0003-6951, 1077-3118},
	url = {http://aip.scitation.org/doi/10.1063/1.1645992},
	doi = {10.1063/1.1645992},
	language = {en},
	number = {6},
	urldate = {2019-05-20},
	journal = {Applied Physics Letters},
	author = {Fujii, T. and Gao, Y. and Sharma, R. and Hu, E. L. and DenBaars, S. P. and Nakamura, S.},
	month = feb,
	year = {2004},
	pages = {855--857}
}

@article{Greiner2007,
	title = {Light {Extraction} from {Organic} {Light} {Emitting} {Diode} {Substrates}: {Simulation} and {Experiment}},
	volume = {46},
	url = {https://doi.org/10.1143\%2Fjjap.46.4125},
	doi = {10.1143/jjap.46.4125},
	number = {7A},
	journal = {Japanese Journal of Applied Physics},
	author = {Greiner, Horst},
	month = jul,
	year = {2007},
	pages = {4125--4137}
}

@article{Whittaker1999,
  title = {Scattering-matrix treatment of patterned multilayer photonic structures},
  author = {Whittaker, D. M. and Culshaw, I. S.},
  journal = {Phys. Rev. B},
  volume = {60},
  issue = {4},
  pages = {2610--2618},
  numpages = {0},
  year = {1999},
  month = {Jul},
  publisher = {American Physical Society},
  doi = {10.1103/PhysRevB.60.2610},
  url = {https://link.aps.org/doi/10.1103/PhysRevB.60.2610}
}

@ARTICLE{Yee1966,
author={ {Kane Yee}},
journal={IEEE Transactions on Antennas and Propagation},
title={Numerical solution of initial boundary value problems involving maxwell's equations in isotropic media},
year={1966},
volume={14},
number={3},
pages={302-307},
doi={10.1109/TAP.1966.1138693},
month={May},}

@article{Moharam1981,
author = {M. G. Moharam and T. K. Gaylord},
journal = {J. Opt. Soc. Am.},
number = {7},
pages = {811--818},
publisher = {OSA},
title = {Rigorous coupled-wave analysis of planar-grating diffraction},
volume = {71},
month = {Jul},
year = {1981},
url = {http://www.osapublishing.org/abstract.cfm?URI=josa-71-7-811},
doi = {10.1364/JOSA.71.000811},
}

@article{Li1996,
author = {Lifeng Li},
journal = {J. Opt. Soc. Am. A},
number = {5},
pages = {1024--1035},
publisher = {OSA},
title = {Formulation and comparison of two recursive matrix algorithms for modeling layered diffraction gratings},
volume = {13},
month = {May},
year = {1996},
url = {http://josaa.osa.org/abstract.cfm?URI=josaa-13-5-1024},
doi = {10.1364/JOSAA.13.001024},
}

@inproceedings{Burger2013,
author = {Sven Burger and Lin Zschiedrich and Jan Pomplun and Frank Schmidt and Bernd Bodermann},
title = {{Fast simulation method for parameter reconstruction in optical metrology}},
volume = {8681},
booktitle = {Metrology, Inspection, and Process Control for Microlithography XXVII},
editor = {Alexander Starikov and Jason P. Cain},
organization = {International Society for Optics and Photonics},
publisher = {SPIE},
pages = {380 -- 386},
year = {2013},
doi = {10.1117/12.2011154},
URL = {https://doi.org/10.1117/12.2011154}
}

@ARTICLE{Santbergen2017,
author={R. {Santbergen} and T. {Meguro} and T. {Suezaki} and G. {Koizumi} and K. {Yamamoto} and M. {Zeman}},
journal={IEEE Journal of Photovoltaics},
title={GenPro4 Optical Model for Solar Cell Simulation and Its Application to Multijunction Solar Cells},
year={2017},
volume={7},
number={3},
pages={919-926},
doi={10.1109/JPHOTOV.2017.2669640},
month={May},}

@article{Eisenlohr2015,
author = {Johannes Eisenlohr and Nico Tucher and Oliver H\"{o}hn and Hubert Hauser and Marius Peters and Peter Kiefel and Jan Christoph Goldschmidt and Benedikt Bl\"{a}si},
journal = {Opt. Express},
number = {11},
pages = {A502--A518},
publisher = {OSA},
title = {Matrix formalism for light propagation and absorption in thick textured optical sheets},
volume = {23},
month = {Jun},
year = {2015},
url = {http://www.opticsexpress.org/abstract.cfm?URI=oe-23-11-A502},
doi = {10.1364/OE.23.00A502},
}

@article{Liu2012,
title = "S4 : A free electromagnetic solver for layered periodic structures",
journal = "Computer Physics Communications",
volume = "183",
number = "10",
pages = "2233 - 2244",
year = "2012",
issn = "0010-4655",
doi = "https://doi.org/10.1016/j.cpc.2012.04.026",
url = "http://www.sciencedirect.com/science/article/pii/S0010465512001658",
author = "Victor Liu and Shanhui Fan"
}

@article{Reinhardt2003,
author = {Reinhardt,Alexandre  and Pastureaud,Thomas  and Ballandras,Sylvain  and Laude,Vincent },
title = {Scattering matrix method for modeling acoustic waves in piezoelectric, fluid, and metallic multilayers},
journal = {Journal of Applied Physics},
volume = {94},
number = {10},
pages = {6923-6931},
year = {2003},
doi = {10.1063/1.1621053},

URL = { 
        https://doi.org/10.1063/1.1621053
    
},
eprint = { 
        https://doi.org/10.1063/1.1621053
    
}

}

@article{Ko1988,
  title = {Matrix method for tunneling in heterostructures: Resonant tunneling in multilayer systems},
  author = {Ko, David Yuk Kei and Inkson, J. C.},
  journal = {Phys. Rev. B},
  volume = {38},
  issue = {14},
  pages = {9945--9951},
  numpages = {0},
  year = {1988},
  month = {Nov},
  publisher = {American Physical Society},
  doi = {10.1103/PhysRevB.38.9945},
  url = {https://link.aps.org/doi/10.1103/PhysRevB.38.9945}
}

@article{Madigan2000,
	title = {Improvement of output coupling efficiency of organic light-emitting diodes by backside substrate modification},
	volume = {76},
	issn = {0003-6951, 1077-3118},
	url = {http://aip.scitation.org/doi/10.1063/1.126124},
	doi = {10.1063/1.126124},
	language = {en},
	number = {13},
	urldate = {2019-05-20},
	journal = {Applied Physics Letters},
	author = {Madigan, C. F. and Lu, M.-H. and Sturm, J. C.},
	month = mar,
	year = {2000},
	pages = {1650--1652}
}

@article{Zhang2016,
  title={High-quality AlN epitaxy on nano-patterned sapphire substrates prepared by nano-imprint lithography},
  author={Zhang, Lisheng and Xu, Fujun and Wang, Jiaming and He, Chenguang and Guo, Weiwei and Wang, Mingxing and Sheng, Bowen and Lu, Lin and Qin, Zhixin and Wang, Xinqiang and others},
  journal={Scientific reports},
  volume={6},
  pages={35934},
  year={2016},
  publisher={Nature Publishing Group}
}

@article{Zhang2013,
	title = {Surface-plasmon-enhanced {GaN}-{LED} based on a multilayered {M}-shaped nano-grating},
	volume = {21},
	issn = {1094-4087},
	url = {https://www.osapublishing.org/oe/abstract.cfm?uri=oe-21-11-13492},
	doi = {10.1364/OE.21.013492},
	language = {en},
	number = {11},
	urldate = {2019-05-20},
	journal = {Optics Express},
	author = {Zhang, Haosu and Zhu, Jun and Zhu, Zhendong and Jin, Yuanhao and Li, Qunqing and Jin, Guofan},
	month = jun,
	year = {2013},
	pages = {13492}
}

@article{Lee2014,
	title = {Enhancement of light-extraction efficiency of organic light-emitting diodes using silica nanoparticles embedded in {TiO}$_2$ matrices},
	volume = {22},
	issn = {1094-4087},
	url = {https://www.osapublishing.org/oe/abstract.cfm?uri=oe-22-S3-A705},
	doi = {10.1364/OE.22.00A705},
	language = {en},
	number = {S3},
	urldate = {2019-05-20},
	journal = {Optics Express},
	author = {Lee, Jooyoung and Kwon, Yun Young and Choi, Eun-Ho and Park, JeongWoo and Yoon, Hong and Kim, Hyunbin},
	month = may,
	year = {2014},
	pages = {A705}
}

@article{Mont2008,
	title = {High-refractive-index {TiO}$_2$-nanoparticle-loaded encapsulants for light-emitting diodes},
	volume = {103},
	issn = {0021-8979, 1089-7550},
	url = {http://aip.scitation.org/doi/10.1063/1.2903484},
	doi = {10.1063/1.2903484},
	language = {en},
	number = {8},
	urldate = {2019-05-20},
	journal = {Journal of Applied Physics},
	author = {Mont, Frank W. and Kim, Jong Kyu and Schubert, Martin F. and Schubert, E. Fred and Siegel, Richard W.},
	month = apr,
	year = {2008},
	pages = {083120}
}

@article{Ponce1997,
	title = {Nitride-based semiconductors for blue and green light-emitting devices},
	volume = {386},
	issn = {0028-0836, 1476-4687},
	url = {http://www.nature.com/articles/386351a0},
	doi = {10.1038/386351a0},
	language = {en},
	number = {6623},
	urldate = {2019-05-20},
	journal = {Nature},
	author = {Ponce, F. A. and Bour, D. P.},
	month = mar,
	year = {1997},
	pages = {351--359}
}

@article{Pomplun2007,
	author = {Pomplun, Jan and Burger, Sven and Zschiedrich, Lin and Schmidt, Frank},
	title = {Adaptive finite element method for simulation of optical nano structures},
	journal = {physica status solidi (b)},
	volume = {244},
	number = {10},
	pages = {3419-3434},
	doi = {10.1002/pssb.200743192},
	url = {https://onlinelibrary.wiley.com/doi/abs/10.1002/pssb.200743192},
	eprint = {https://onlinelibrary.wiley.com/doi/pdf/10.1002/pssb.200743192},
	year = {2007}
}

@article{Martens2016,
author = {Martens,M.  and Kuhn,C.  and Ziffer,E.  and Simoneit,T.  and Kueller,V.  and Knauer,A.  and Rass,J.  and Wernicke,T.  and Einfeldt,S.  and Weyers,M.  and Kneissl,M. },
title = {Low absorption loss p-{AlGaN} superlattice cladding layer for current-injection deep ultraviolet laser diodes},
journal = {Applied Physics Letters},
volume = {108},
number = {15},
pages = {151108},
year = {2016},
doi = {10.1063/1.4947102},

URL = { 
        https://doi.org/10.1063/1.4947102
    
},
eprint = { 
        https://doi.org/10.1063/1.4947102
    
}

}

@article{Liang2018,
author = {Liang,Y.-H.  and Towe,E. },
title = {Progress in efficient doping of high aluminum-containing group {III}-nitrides},
journal = {Applied Physics Reviews},
volume = {5},
number = {1},
pages = {011107},
year = {2018},
doi = {10.1063/1.5009349},

URL = { 
        https://doi.org/10.1063/1.5009349
    
},
eprint = { 
        https://doi.org/10.1063/1.5009349
    
}

}

@book{Lavrinenko2014,
title ={Numerical Methods in Photonics},
publisher = {CRC Press},
year = {2014},
author = {Lavrinenko, A. V. and Laegsgaard, J. and Gregersen, N. and Schmidt, F. and Soendergaard, T.}
}

@ARTICLE{Ooi2018,
author={Y. K. {Ooi} and J. {Zhang}},
journal={IEEE Photonics Journal},
title={Light Extraction Efficiency Analysis of Flip-Chip Ultraviolet Light-Emitting Diodes With Patterned Sapphire Substrate},
year={2018},
volume={10},
number={4},
pages={1-13},
doi={10.1109/JPHOT.2018.2847226},
month={Aug},}

@article{Lee2017,
author = {Lee,Donghyun  and Lee,Jong Won  and Jang,Jeonghwan  and Shin,In-Su  and Jin,Lu  and Park,Jun Hyuk  and Kim,Jungsub  and Lee,Jinsub  and Noh,Hye-Seok  and Kim,Yong-Il  and Park,Youngsoo  and Lee,Gun-Do  and Park,Yongjo  and Kim,Jong Kyu  and Yoon,Euijoon },
title = {Improved performance of {AlGaN}-based deep ultraviolet light-emitting diodes with nano-patterned {AlN}/sapphire substrates},
journal = {Applied Physics Letters},
volume = {110},
number = {19},
pages = {191103},
year = {2017},
doi = {10.1063/1.4983283},

URL = { 
        https://doi.org/10.1063/1.4983283
    
},
eprint = { 
        https://doi.org/10.1063/1.4983283
    
}

}

@article{Dong2013,
	title = {282-nm {AlGaN}-based deep ultraviolet light-emitting diodes with improved performance on nano-patterned sapphire substrates},
	volume = {102},
	issn = {0003-6951},
	url = {https://aip.scitation.org/doi/full/10.1063/1.4812237},
	doi = {10.1063/1.4812237},
	number = {24},
	urldate = {2019-09-18},
	journal = {Applied Physics Letters},
	author = {Dong, Peng and Yan, Jianchang and Wang, Junxi and Zhang, Yun and Geng, Chong and Wei, Tongbo and Cong, Peipei and Zhang, Yiyun and Zeng, Jianping and Tian, Yingdong and Sun, Lili and Yan, Qingfeng and Li, Jinmin and Fan, Shunfei and Qin, Zhixin},
	month = jun,
	year = {2013},
	pages = {241113}
}

@article{Takano2017,
	title = {Deep-ultraviolet light-emitting diodes with external quantum efficiency higher than 20\% at 275 nm achieved by improving light-extraction efficiency},
	volume = {10},
	issn = {1882-0786},
	url = {https://iopscience.iop.org/article/10.7567/APEX.10.031002/meta},
	doi = {10.7567/APEX.10.031002},
	language = {en},
	number = {3},
	urldate = {2019-09-17},
	journal = {Applied Physics Express},
	author = {Takano, Takayoshi and Mino, Takuya and Sakai, Jun and Noguchi, Norimichi and Tsubaki, Kenji and Hirayama, Hideki},
	month = feb,
	year = {2017},
	pages = {031002}
}

@article{Inoue2017,
	title = {150 {mW} deep-ultraviolet light-emitting diodes with large-area {AlN} nanophotonic light-extraction structure emitting at 265 nm},
	volume = {110},
	issn = {0003-6951},
	url = {https://aip.scitation.org/doi/10.1063/1.4978855},
	doi = {10.1063/1.4978855},
	number = {14},
	urldate = {2019-09-18},
	journal = {Applied Physics Letters},
	author = {Inoue, Shin-Ichiro and Tamari, Naoki and Taniguchi, Manabu},
	month = apr,
	year = {2017},
	pages = {141106}
}

@article{Hagedorn2016,
author = {Hagedorn, Sylvia and Knauer, Arne and Mogilatenko, Anna and Richter, Eberhard and Weyers, Markus},
title = {AlN growth on nano-patterned sapphire: A route for cost efficient pseudo substrates for deep UV LEDs},
journal = {physica status solidi (a)},
volume = {213},
number = {12},
pages = {3178-3185},
doi = {10.1002/pssa.201600218},
url = {https://onlinelibrary.wiley.com/doi/abs/10.1002/pssa.201600218},
eprint = {https://onlinelibrary.wiley.com/doi/pdf/10.1002/pssa.201600218},
year = {2016}
}

@article{Pastrnak1966,
author = {Pastrňák, J. and Roskovcová, L.},
title = {Refraction Index Measurements on {AlN} Single Crystals},
journal = {physica status solidi (b)},
volume = {14},
number = {1},
pages = {K5-K8},
doi = {10.1002/pssb.19660140127},
url = {https://onlinelibrary.wiley.com/doi/abs/10.1002/pssb.19660140127},
eprint = {https://onlinelibrary.wiley.com/doi/pdf/10.1002/pssb.19660140127},
year = {1966}
}

@incollection{Dodge1986,
author = {M. J. Dodge},
title = {Refractive Index},
booktitle = {Handbook of Laser Science and Technology Supplement 2: Optical Materials},
editor = {Marvin J. Weber},
publisher = {CRC Press, Boca Raton},
year = {1986},
pages = {30}
}
}

\end{document}